\documentclass[11pt]{article}
\usepackage{epsfig}
\psfigdriver{dvips}
\newtheorem{theo}{Theorem}

\newtheorem{prop}[theo]{Proposition}
\newtheorem{cor}[theo]{Corollary}

\input{epsf}

\begin{document}

\def\C{{\rm\kern.24em \vrule width.02em height1.4ex
depth-.05ex\kern-.26em C}}
\def\N{{\rm I\hspace{-0.4ex}N}} 
\def\R{{\rm I\hspace{-0.4ex}R}}
\def\rit{{\rm I\hspace{-0.4ex}R}}
\def\grad{\hbox{ grad }}
\def\curl{\hbox{curl }}
\def\div{\hbox{ div }}
\def\supp{{\rm supp}\:}
\def\Lip{{\rm Lip}}
\def\sign{{\rm sign}\:}
\def\dist{{\rm dist}}
\def\diam{{\rm diam}\:}
\def\const{{\rm const.}\:}
\def\meas{{\rm meas}\:}
\def\oOmega{\overline{\Omega}}
\def\eps{\varepsilon}
\def\be{\begin{equation}}
\def\ee{\end{equation}}
\def\beq{\begin{equation}}
\def\eeq{\end{equation}}
\def\ds{\displaystyle}
\def\ts{\textstyle}
\def\qed{\rule{1ex}{1ex}}   

\title{\bf  On the symmetry and uniqueness of solutions of the Ginzburg-Landau equations for small domains}
\author{A.Aftalion\footnote{Laboratoire d'Analyse Num\'erique, B.C.187, Universit\'e Pierre et Marie Curie, 4 Place Jussieu, 75252 Paris cedex 05,  France. }$\ $ and E.N.Dancer\footnote{School of Mathematics and Statistics, University of Sydney, N.S.W.2006, Australia, partially supported by ARC.}}
\maketitle

\begin{abstract}
In this paper, we study the Ginzburg-Landau equations for a two dimensional domain which has small size. We prove that if the domain is small, then the solution has no zero, that is no vortex. More precisely, we show that the order parameter $\Psi$ is almost constant. Additionnally, we obtain that if the domain is a disc of small radius, then any non normal solution is symmetric and unique. Then, in the case of a slab, that is a one dimensional domain, we use the same method to derive that solutions are symmetric. The proofs use a priori estimates and the Poincar\'e inequality.
\end{abstract}

\section{Introduction and main results}
\setcounter{equation}{0}

In this paper, we study the properties of a superconducting cylinder
 submitted to an exterior magnetic field $\hbox{\bf H$_0$}$ parallel to
 the axis of the cylinder. According to the Ginzburg-Landau theory of superconductivity, the sample is in a state that minimizes the following energy:

\beq\label{engl}
{E}_{\kappa}(\Psi,\hbox{\bf A})=\int_{\Omega}|({1 \over \kappa} \nabla -i\hbox
{\bf A})\Psi |^2+
{1\over2}(|\Psi|^2-1)^2+(\curl\hbox{\bf A}-\hbox{\bf H$_0$})^2 \ d\Omega.
\eeq      
Here $\Omega$ is a simply connected domain in $\rit^2$ with characteristic length $d$.
 As usual, $\kappa$ is the Ginzburg-Landau parameter that characterizes the type
 of superconductor, $\hbox{\bf A}(x,y)$ is the vector potential, so that $\curl \hbox{\bf A}$ is the magnetic field and $\Psi (x,y) $ is the order parameter. Because $\hbox{\bf H$_0$}=H_0\hbox{\bf e$_z$}$ is along the $z$ axis, we can assume without loss of generality that $\hbox{\bf A}=(A_1,A_2,0)$. For a detailed description of the model, one may refer to \cite{du} or \cite{T}.

 The minimization process yields the classical Ginzburg-Landau equations (see \cite{du} for more details):

$$\left\{\begin{array}{ll}
({1 \over \kappa} \nabla-i\hbox{\bf A})^2 \Psi=\Psi(|\Psi|^2-1) \quad \hbox{in}
 \quad \Omega,\\

-\curl \curl \hbox{\bf A}={i \over {2 \kappa}} (\Psi^* \nabla \Psi-\Psi \nabla \Psi^*)
+\hbox{\bf A}|\Psi|^2 \quad \hbox{in}\quad \Omega,\\

{{\partial \Psi} \over {\partial n}}-i\Psi \hbox{\bf A}\cdot \hbox{\bf n}=0 \quad \hbox{on}\quad  \partial \Omega,\\

\curl \hbox{\bf A}\times \hbox{\bf n} =\hbox{\bf H}_0\times \hbox{\bf n}\quad \hbox{on}\quad  \partial \Omega.
\end{array}\right.\eqno{(GL)}$$
 It is common to use the gauge where $\div \hbox{\bf A} =0$ and $\hbox{\bf A} \cdot \hbox{\bf n}=0$ on $\partial \Omega$.
 Notice that this system has a special set of solutions with $\Psi \equiv 0$ and $\curl \hbox{\bf A} \equiv \hbox{\bf H}_0$, called {\em normal solutions}. They correspond to a case where superconductivity is destroyed. In the following, we will be concerned with solutions which are not in this set. 

In this problem, there are 3 variable parameters: $\kappa$, $H_0$ and $d$ the characteristic size of $\Omega$. According to the values of these parameters, the properties of solutions (existence of non normal solutions, number, symmetry) change. It has been the purpose of various authors to study the properties of solutions in some asymptotic limits: for instance, Giorgi and Phillips \cite{GP} have proved that for $\Omega$ and $\kappa$ fixed and $H_0$ large enough, then the only solution is the normal solution. In the case where $d$ is fixed and $\kappa$ is large, Sandier and Serfaty \cite{SS} have  proved  that the solutions can have a lot of vortices for a certain range of magnetic field. There have been many other contributions in other asymptotic regimes.

    In this paper, we are interested in the case where $d$, the characteristic length of $\Omega$ is small, in particular compared to $1/\kappa$.  In this setting, it is expected that $\Psi$ is almost constant and has no zero, that is no vortices, since a vortex is of size $1/\kappa$. This result is often used in the physics literature, but we did not know any rigorous proof. This is what we show:

\begin{theo}\label{main}
Assume that $D$ is a fixed simply connected bounded domain and let 
 $\Omega=dD$.
For any $d_0>0$, there exists $d_1>0$, such that if $d<\min (d_0, d_1 / \kappa)$, for any $H_0=H_0(d,\kappa)$, then any solution of $(GL)$ which is not a normal solution is such that $\Psi$ has no zero. Moreover, $|\Psi / |\Psi|_\infty -1|\leq C\kappa d$.
\end{theo} 
We let $H_0$ be a function of $d$ and $\kappa$ because it is expected that a superconducting solution exists up to fields $H_0$ of the order of $C/d$ as we will explain in more details later.
 Additionally, we prove a symmetry and uniqueness result when the domain is a small disc:
\begin{theo}\label{mainsym}
Let $\Omega$ be a disc of radius $d$. There exist constants $d_0$ and $d_1$, such that if $d<\min (d_0, d_1 / \kappa)$, for any $H_0=H_0(d,\kappa)$, then any solution of $(GL)$ is radially symmetric, that is $\Psi (x,y)=\Psi (r)$ and $\hbox{\bf A} (x,y)=A(r)\hbox{\bf e}_\theta$, where $r =\sqrt { x^2+y^2}$.
\end{theo}
\begin{theo}\label{mainuniq}
Under the hypotheses of Theorem \ref{mainsym},
 if there exists a non normal solution $(\Psi,\hbox{\bf A})$ of  $(GL)$, then this solution is unique up to multiplication of $\Psi$ by a constant of modulus 1.
\end{theo}
\begin{cor}\label{coruniq}
Under the hypotheses of Theorem \ref{mainsym}, there exists $H_*=H_*(d,\kappa)$, with $\lim_{d\to 0} d H_*(d,\kappa)= 2\sqrt 2$ and for $H_0<H_*$, there exists a unique non normal solution of $(GL)$, while for $H_0\geq H_*$, the only solutions are the normal solutions.
\end{cor}

We make a change of variables $x'=x /d$, $y'=y/d$ so that the new variable lies in a domain of unit size $D$.
 We also define $\hbox{\bf B}=\kappa d \hbox{\bf A}$ and $ h_0= \kappa d^2 H_0$. In the following, we will assume additionnally that 
\beq\label{div}
\div\hbox{\bf B}=0 \quad \hbox{in} \quad \Omega \quad \hbox{and} \quad \hbox{\bf B}\cdot \hbox{\bf n}=0 \quad \hbox{on} \quad \partial \Omega.
\eeq
  The equations then become

$$\left\{\begin{array}{ll}
( \nabla-i \hbox{\bf B})^2 \psi=\kappa^2 d^2 \psi(|\psi|^2-1) \quad \hbox{in}
 \quad D,\\
 \Delta \hbox{\bf B}= d^2 \bigl ({{i} \over {2} } (\psi^* \nabla \psi-\psi \nabla
  \psi^*)
  + \hbox{\bf B}|\psi|^2\bigr )  \quad \hbox{in}\quad D,\\
 
  {{\partial \psi} \over {\partial n}}=0 \quad \hbox{on}\quad  \partial D,
  \\
  \curl \hbox{\bf B}\times \hbox{\bf n}=\hbox{\bf h}_0\times \hbox{\bf n}\quad \hbox{on}\quad  \partial D.
\end{array}\right.\eqno{(GL_d)}$$
Note that another way of writing the equation for $\hbox{\bf B}$ is 
\beq\label{eqB}
\Delta \hbox{\bf B}=-d^2  (i\psi , \nabla \psi -i \hbox{\bf B}\psi),
\eeq
where $(\ .\ ,\ .\ )$ is the real part of the scalar product in $\C$.

We allow $h_0$ to vary with $d$ and $\kappa$ but we will prove that  if there exists a solution, then in fact $ h_0$ is bounded, that is $H_0$ is bounded by $C/\kappa d^2$. 

The proof consists in obtaining a priori estimates for the solutions $(\psi,\hbox{\bf B})$. This is done in Section 2.
 In Section 3, we use these a priori estimates and the Poincar\'e inequality to derive that $\psi$ is nearly constant, hence has no zero. This will prove Theorem \ref{main}.
 Then, in Section 4, we define the functions 
$$\tilde\psi(x,y)=\psi(-x,y), \quad \tilde{\hbox{\bf B}}(x,y)=\left(\begin{array}{ll}
\ \ -B_1(-x,y)\\
B_2(-x,y)\end{array}\right)
,$$
\beq\label{wz}
w(x,y)={1 \over {|\psi|_\infty}}(\psi(x,y)-\tilde\psi(x,y)), \quad \hbox{\bf z}(x)=\hbox{\bf  B}(x,y)-\tilde{\hbox{\bf  B}}(x,y),
\eeq
which  satisfy elliptic PDE's with small right hand side terms. Then we use that $\psi$ is nearly constant and  hence we get that $w$ and $\hbox{\bf z}$ are identically zero. This will prove Theorem \ref{mainsym}.
 In Section 5, we obtain the uniqueness result proving that any solution is necessary a local minimizer of the energy.
 Finally in Section 6, we show how our proof can be adapted to the one dimensional case to provide a symmetry result in this setting and in Section 7 we summarize our results in terms of bifurcation curves and prove Corollary \ref{coruniq}.

\noindent
{\bf Remark:} If $\Omega$ is simply connected and symmetric in the $y$ direction, then our proof also gives that $\psi$ and $\hbox{\bf  B}$ are symmetric, in the sense that $\tilde{\psi}=\psi$ and $\tilde{\hbox{\bf  B}}=\hbox{\bf  B}$.

\hfill

We note that in the case with no magnetic field, Bethuel, Brezis and Helein \cite{BBH} have proved the symmetry of solutions when the domain is small using the Poincar\'e inequality. 
The radial symmetry of local minimizers has also been studied in \cite{M} for a different system with no magnetic field. Here the proof is more involved since we have to deal with the system with applied magnetic field.

Let us point out that the uniqueness proof of Section 5 is inspired by \cite{ABG} where it is obtained that any radial solution with a giant vortex of degree $N$ at the origin is a local minimizer of the energy.

\section{A priori estimates}
\setcounter{equation}{0}

\begin{prop}\label{aprioriest}
 Fix $p>1$. Assume that $(psi,\hbox{\bf B})$ is a non normal solution of $(GL_d)$ and (\ref{div}).
 For all constants $d_0$ and $d_1$, if $d<\min (d_0,$ $ d_1 / \kappa)$, then
 $\psi$ and $\hbox{\bf B}$ are bounded in $W^{2,p}(D)$ by constants independent of $d$ and $\kappa$. Moreover, for fixed $\kappa$,
\beq\label{limd0}
\lim_{d\to 0} {\hbox{ h}_0}=0\quad \hbox{and}\quad\lim_{d\to 0}\| {\hbox{\bf  B}} \|_{W^{2,p}(D)}=0.
\eeq
\end{prop}
The goal of this section consists in proving Proposition \ref{aprioriest}.

\subsection{First estimates for $\psi$}

First recall from Du-Gunzburger-Peterson \cite{du} that
\beq\label{psi1}
|\psi | \leq 1 \quad \hbox{a.e.}
\eeq
Next we define
\beq\label{upi}
u(x,y)=|\psi (x,y)|^2 \quad \hbox{and} \quad {\bf \Pi}=\nabla-i  \hbox{\bf B}
.
\eeq
One can compute easily that 
\beq\label{equ}
{1 \over 2}\Delta u +\kappa ^2 d^2 u(1-u)=|{\bf \Pi} \psi |^2.
\eeq
Integrate (\ref{equ}) in $D$, use the boundary condition $\partial \psi / \partial n =0$  and get the key estimate for $\psi$:
\beq\label{ineqpi}
\| {\bf \Pi} \psi \| _{L^2(D)} \leq \kappa d \|\psi\| _{L^2(D)}.
\eeq

\subsection{Estimates for $\hbox{\bf B}$}

We decompose $\hbox{\bf B}=h_0 \hbox{\bf b}_0+\hbox{\bf b}_r$
 where $\hbox{\bf b}_0$ is chosen such that 
$$
\curl \curl \hbox{\bf b}_0 =0 \ \hbox{and} \ \div 
\hbox{\bf b}_0 =0 \ \hbox{in} \ D, \
\curl \hbox{\bf b}_0  =\hbox{\bf e}_z \ \hbox{and} \ \hbox{\bf b}_0 \cdot \hbox{\bf n}=0 \  \hbox{on} \  \partial D,
$$
and $\hbox{\bf b}_r=\hbox{\bf B}-h_0 \hbox{\bf b}_0$.

A particular choice for $\hbox{\bf b}_0$ when $\Omega$ is a disc
 is $\hbox{\bf b}_0 = {1 \over 2} (-y,x,0)$. Note that in any case $\hbox{\bf b}_0$ and $\hbox{\bf b}_r$ are vectors in the $x-y$ plane and their $curl$ are in the $z$ direction.
Making the difference between the equations for $\hbox{\bf B}$ and $\hbox{\bf b}_0$, we obtain the equation for $\hbox{\bf b}_r$ 
\begin{eqnarray}
\label{eqbr}
&&\curl\curl \hbox{\bf b}_r = d^2 (i \psi , {\bf \Pi} \psi) \quad \hbox{and}\quad \div \hbox{\bf b}_r =0 \quad \hbox{in}\ D, \\
\label{eqbrbound}
&& \curl \hbox{\bf b}_r = 0 \quad \hbox{and} \quad \hbox{\bf b}_r\cdot \hbox{\bf n} = 0 \quad \hbox{on}\quad \partial D.
\end{eqnarray}
Since $ \div 
\hbox{\bf b}_r =0$ and $\hbox{\bf b}_r\cdot \hbox{\bf n} = 0$ on $ \partial D$,
 then $\|\curl \hbox{\bf b}_r\|_{L^p(D)}$ is a norm in $W^{1,p}(D)$, that is
\beq\label{normcurl}
\| \hbox{\bf b}_r\|_{W^{1,p}(D)}\leq C \|\curl \hbox{\bf b}_r\|_{L^{p}(D)}.
\eeq
 Now Cauchy-Schwarz inequality and (\ref{eqbr}) imply that 
 \beq\label{ineqbr}
 | \curl \curl \hbox{\bf b}_r |_2 \leq d^2 | {\bf \Pi }\psi|_2.
 \eeq
Note that $|\ .\ |_2$ is the 2-norm for vectors in $\rit^2$. From (\ref{ineqbr}),  we get the $W^{2,2}$ bound for $\hbox{\bf b}_r$. 
Thus, because of (\ref{ineqpi}), (\ref{ineqbr}) and the boundary condition (\ref{eqbrbound}), it follows from the Poincar\'e inequality that 
\beq\label{curlest2}
\|\curl \hbox{\bf b}_r\|_{W^{1,2}(D)}\leq C \kappa d^3.
\eeq
We gather (\ref{normcurl}) and (\ref{curlest2}) to obtain the key estimate for $\hbox{\bf b}_r$:
\beq\label{estbr}
\|\hbox{\bf b}_r \|_{W^{2,2}(D)} \leq C \kappa d^3,
\eeq
where $C$ is independent of $d$ and $\kappa$.
In order to get the bounds for $\hbox{\bf B}$, we only need to prove
 that $\lim _{d\to 0}h_0=0 $.
  Let us assume that $h_0$ is bounded below by some constant $m$. Then
 we use an estimate by Giorgi and Phillips \cite{GP} (Lemma 2.8 p.349): there exists
 $C_2$ (which depends on $m$) such that

\beq\label{eqgiorgi}
C_2 h_0 \int_D |\psi|^2 \leq \int_D | \nabla \psi -i h_0  \hbox{\bf b}_0 \psi |^2
.
\eeq
In order to bound the right hand side of (\ref{eqgiorgi}), we write $h_0 \hbox{\bf b}_0= \hbox{\bf B}-\hbox{\bf b}_r$ and use
 (\ref{ineqpi}) and (\ref{estbr}) to get
 \beq\label{h0bdd}
 {(C_2  h_0 )}^{1 /2}(\int_D |\psi |^2)^{1 /2} \leq \kappa d (\int_D |\psi |^2)^{1 /2} (1+C_1 d^2).
 \eeq

So we obtain a contradiction with the lower bound for $h_0$ when $d$ is small enough.
 Hence $\hbox{\bf B}$ is bounded in 
 $W^{2,2}$, independently of $d$ and $\kappa$. 

Once we have bounded $\hbox{\bf B}$, the equation for $\psi$ can be thought of
 as a linear elliptic equation with bounded coefficients, so that the classical elliptic estimates imply bounds for $\psi$.

Finally, in order to get the $W^{2,p}$ bound for $\hbox{\bf B}$, that is for $\hbox{\bf b}_r$, we bootstrap equation (\ref{eqbr}) (see \cite{GT}). Thus we obtain that 
\beq\label{curlest3}
\|\curl \hbox{\bf b}_r\|_{W^{2,p}(D)}\leq C  d^2.
\eeq
and (\ref{limd0}) is true for $\hbox{\bf B}$.

\section{$\psi$ is nearly constant}
\setcounter{equation}{0}
\begin{prop}\label{psiconst}
 For all $d_0>0$, there exists $d_1>0$, such that
 if $d< \min (d_0,d_1 /\kappa)$, then $\|\nabla \psi / |\psi|_\infty \|_\infty$ and $|\psi / |\psi| _{\infty} -1|$
 are  small, thus $\psi$ has no zero.
\end{prop}
We define $\phi=\psi / | \psi|_\infty$, then the equation for $\phi$ can be written as
$$\Delta \phi-2i  \hbox{\bf B} \cdot \nabla \phi=\kappa^2 d^2 \phi(|\psi|_\infty^2 |\phi|^2-1)+|\hbox{\bf B}|^2 \phi$$
and for $p<2$
\beq\label{psiest}
|\hbox{\bf B}\cdot \nabla \phi|_{p}\leq |\hbox{\bf B}\cdot {\bf \Pi} \phi|_{p}+|\hbox{\bf B}^2\phi|_{p}\leq |{1\over {|\psi|_\infty}} {\bf \Pi} \psi|_2|\hbox{\bf B}|_{{2p}\over{2-p}}+|\hbox{\bf B}^2|_p.\eeq
 Since $|\phi |\leq 1$,
 and because of (\ref{ineqpi}) and the previous estimates for $\hbox{\bf B} $,
 we see that the right hand side of (\ref{psiest}) is small when $d$ is small.
 Hence we go back to the equation for $\phi$ and the Agmon Douglis Nirenberg estimates imply a bound for $\phi$ in $W^{2,p}$ for any $p<2$. Hence by Sobolev embedding, $\nabla \phi$ is bounded in $L^p$ for any $p$ and going back to the equation for $\phi$, we derive a bound for $\phi$ in $W^{2,p}$ for any $p$,
hence in $C^1$. In particular, $\nabla \phi$ is equicontinuous. Now, when $d$ tends to 0, we know that $\|\hbox{\bf B}\|_\infty$ tends to 0, so if we multiply the equation for $\phi$ by $\phi^*$ and integrate, we find that $\|\nabla  \phi \| _{L^2(D)}$ is small. Thus, since $\nabla \phi$ is equicontinuous, we see that $\|\nabla \phi \|_\infty$ tends to 0 with $d$,
 which proves the first part of Proposition \ref{psiconst}.

In fact we can get a precise estimate for the smallness of $|\psi / |\psi| _{\infty} -1|$.
Take the equation (\ref{equ}), multiply by $u$ and integrate to get
\beq\label{ineqgradu}
\int_D |\nabla u|^2 \leq \kappa^2 d^2 \int _D u^2 (1-u).
\eeq
Next we define $v=u/| \psi|^2_
\infty$. Then (\ref{ineqgradu}) implies that 
$$\int_D |\nabla v|^2 \leq \kappa^2 d^2 |D|.$$
We call $v_{mean}$ the mean value of $v$ in $D$. The previous estimate and
 the Poincar\'e inequality yield
\beq \label{umean}
\|v -v_{mean} \|_{L^2(D)} \leq C_0 \kappa d.
\eeq
Let us call 
 $r$ an upper bound for $\nabla v$, independent of $d$ and $\kappa$. We have seen in the first part of the proof that it exists. 

Let us now prove that 
\beq\label{vmean}
|v(x) -v_{mean}| \leq \alpha \quad \forall x\in D, \quad \hbox{where} \quad \alpha^2=16C_0 r \kappa d/ \sqrt\pi,
\eeq
where $C_0$ comes from (\ref{umean}).
Assume that (\ref{vmean}) is not true. Then 
there exists a point $x_0$ where this does not hold. Set $\eta=\alpha/ 4r$.
 Note that when $\kappa d$ is small, then $\eta$ is small.
 If $\dist (x_0,\partial D)
  \leq 2 \eta$, 
  then for $\kappa d$ small enough, there always exists a point $x_1$ in $D$ with $\dist (x_1,\partial D)
  \geq 2 \eta$ and $\dist (x_1,x_0)
  \leq 2 \eta$.  If $\dist (x_0,\partial D)
  \geq 2 \eta$ then we set $x_1=x_0$. Now we have $|v(x)-v_{mean}|>\alpha/4$ for $|x-x_1|\leq
  \eta$. Define $D_1=D(x_1, \eta)$, then
  $$\|v -v_{mean} \|_{L^2(D_1)} > \eta\sqrt \pi \alpha /4.$$
This provides a contradiction with the definition of $\alpha$ and (\ref{umean}).

Now recall that $v$ is necessarily 1 for at least one point in $D$. So we use (\ref{vmean}) to get
$$|\psi(x) / |\psi| _{\infty} -1|\leq C\kappa d \quad \forall x \in D.$$
This implies in particular that for small $\kappa d$, $\psi$ is never equal to zero.

\section{Radial symmetry of the solutions in the case of a disc}
\setcounter{equation}{0}
In this Section, we assume that $\Omega$ is a disc of radius $d$. We are going to use the previous a priori estimates to obtain radial symmetry. 
 We define the functions $w$ and $\hbox{\bf z}$ as in (\ref{wz}). They satisfy the following equations
\begin{eqnarray}
&&\Delta w -2 i \hbox{\bf B}\cdot \nabla w -{{2 i}\over{|\psi|_\infty}} \hbox{\bf z}\cdot \nabla \tilde{\psi}=\kappa^2 d^2 w(|\psi|^2 +\tilde{\psi}(\psi+\tilde{\psi})-1) \nonumber\\
&&\hspace{4cm}+w |\hbox{\bf B}|^2+{\tilde{\psi}\over{|\psi|_\infty}}\hbox{\bf z}\cdot(\hbox{\bf B}+\tilde{\hbox{\bf B}}) \quad \hbox{in}\quad   D,\label{eqw} \\ \label{bcw}
&&\displaystyle{{\partial w} \over {\partial n}}=0 \quad \hbox{on}\quad  \partial D,\\
&&\curl \curl \hbox{\bf z}=d^2 \bigl [ (i|\psi|_\infty w,{\bf \Pi}\psi)+(i\tilde{\psi},|\psi|_\infty \nabla w-i\hbox{\bf B}|\psi|_\infty w-i\tilde{\psi}\hbox{\bf z}) \bigr ]\nonumber \\
&&\hspace{6cm}\hbox{and} \quad \div\hbox{\bf z}=0 \quad \hbox{in}\quad  D,\label{eqz}\\ \label{bcz}
&&\hbox{\bf z}\cdot \hbox{\bf n}=0\quad\hbox{and}\quad \curl \hbox{\bf z}\times \hbox{\bf n}=0 \quad \hbox{on}\quad  \partial D.
\end{eqnarray}
\begin{prop}\label{wzzero}
There are constants $d_0$ and $d_1$ such that, if $d<\min (d_0, $ $d_1/ \kappa)$, then $w\equiv \hbox{\bf z}\equiv 0$.
\end{prop}

Let us multiply (\ref{eqw}) by $w$ and integrate. The first term only gives an integral of the gradient of $w$ because of the boundary condition (\ref{bcw}). We recall that $|\psi|\leq 1$ and $|w|\leq 2$ to obtain
\begin{eqnarray*}
 \int_D|\nabla w|^2 &\leq& 4\|\hbox{\bf B}\|_\infty \int_D|\nabla w|+(4\kappa^2d^2+\|\hbox{\bf B}\|_\infty^2)\int_D w^2\\
& &+4(\|\nabla{\psi\over{\|\psi\|_\infty}}\|_\infty+\|\hbox{\bf B}\|_\infty)\int_D|\hbox{\bf z}|.
\end{eqnarray*}
Then we use our previous estimates for $\psi$ and $\hbox{\bf B}$ in Proposition \ref{aprioriest} and \ref{psiconst}. In particular we recall that $\|\hbox{\bf B}\|_\infty$ and $\|{1 \over{|\psi|_\infty}}\nabla \psi\|_\infty$ are small when $d$ is small to get: for all $\eps>0$, if $d$ is small enough, then
$$\|\nabla w\|_2 \leq \eps \|w\|_2+\eps \|\hbox{\bf z}\|_2.$$
Because of the definition of $w$, one can see easily that $w$ has mean value zero. We can use the Poincar\'e inequality to get
\beq\label{wsmall}
\mu \|w\|_2 \leq \eps \|w\|_2+\eps \|\hbox{\bf z}\|_2,
\eeq
where $\mu>0$ is independent of $d$ and $\kappa$.
 Thus
\beq\label{wlessz}
\|w\|_2 \leq {\eps \over{\mu-\eps}}\|\hbox{\bf z}\|_2\leq \|\hbox{\bf z}\|_2,
\eeq
provided $\eps/ (\mu-\eps)$ is less then 1.

Similarly, we multiply equation (\ref{eqz}) by $\hbox{\bf z}$, integrate
 and use that the boundary condition implies $\curl \hbox{\bf z}=0$ on $\partial D$ to get
$$ \int_D |\curl \hbox{\bf z}|^2_2\leq d^2 \int_D |\hbox{\bf z}|_2 | \Pi\psi+\nabla w+\hbox{\bf B}+\hbox{\bf z}|_2].$$
We can apply the previous estimates which give bounds on $\psi$ and $\hbox{\bf B}$ to derive
\beq\label{zsmall}
\|\nabla \hbox{\bf z}\|_2 \leq \eps \|\hbox{\bf z}\|_2.\eeq
 Then we write  $\hbox{\bf z}=\hbox{\bf z}_m+\hbox{\bf z}_t$ where $\hbox{\bf z}_m$ is the mean value of 
 $\hbox{\bf z}$ (so $\hbox{\bf z}_m$ is a vector) and $\hbox{\bf z}_t$ has mean value zero. Again we use the Poincar\'e inequality for $\hbox{\bf z}_t$ and equation (\ref{zsmall}) to get
\beq\label{zest}
\|\hbox{\bf z}_t\|_2\leq {{ \eps} \over{\mu-\eps}}\|\hbox{\bf z}_m\|_2\quad
\hbox{and}\quad \|w\|_2 \leq {{\eps \mu} \over{(\mu-\eps)^2}}\|\hbox{\bf z}_m\|_2.
\eeq
Now we integrate (\ref{eqz}) over $D$,  use the boundary conditions which imply that the $\curl \curl$ term contribution vanishes and we get
\beq\label{finalest}
\int_D {{|\tilde{\psi}|^2}\over{|\psi|_\infty^2}}\hbox{\bf z}=\int_D (iw,{\bf \Pi}\psi)+(i\tilde{\psi}, \nabla w-i\hbox{\bf B} w).
\eeq
This yields
$$|\hbox{\bf z}_m| \int_D {{|{\psi}|^2}\over{|\psi|_\infty^2}}\leq\|w\|_2\|{\bf \Pi}\psi\|_2+\|\psi\|_2\|\nabla w-i\hbox{\bf B} w\|_2+\int_D |\hbox{\bf z}_t|\leq C\eps|\hbox{\bf z}_m|.$$
The last inequality comes from (\ref{zest}) and the bounds for $\psi$ and $\hbox{\bf B}$ in Section 3.
If $\hbox{\bf z}_m$ is not zero, the boundedness of ${\bf \Pi}\psi$, $\hbox{\bf B}$ and (\ref{zest}) imply that 
$$\int_D |\psi|^2 / |\psi|_\infty^2$$
 is small. This provides a contradiction with Proposition \ref{psiconst}, where we have proved that $|\psi|/|\psi|_\infty$ is nearly 1 for small $d$. 

\hfill

Let us complete the proof of Theorem \ref{mainsym}. Indeed, once we know that $w$ and $\hbox{\bf z}$ are identically zero, then it implies as in \cite{GNN} that $\psi $ and $|\hbox{\bf B}|$ are radial, where $|\ .\ |$ is the modulus. Thus, we see that either $\hbox{\bf B}$ vanishes everywhere on the circle $|(x,y)|=r$ and there is nothing to prove, or nowhere on this circle. In this case, we can argue that $\hbox{\bf B}(r,\theta) $ is in the direction of ${\bf e}_\theta=(-\sin\theta, \cos \theta)$. Indeed, since $\hbox{\bf z}=0$, it implies in particular that $B_1(r,0)=0$, hence $\hbox{\bf B}(r,0) $ is in the direction of $(0,1)$. Then we use a reflection in the axis $T$ in the direction of $(\cos \theta , \sin \theta )$ and derive that the corresponding $w$ and $\hbox{\bf z}$ are zero. This implies that $\hbox{\bf B}(r,\theta)\cdot{\bf e}_r=0 $. Since $B_1(r,0)=0$, a continuity argument implies that $\hbox{\bf B}(r,\theta)=B(r) {\bf e}_\theta$, as we had claimed.

Let us show that one can choose $\psi$ to be real.
Since $\psi$ is radial and has no zero, we can write $\psi=f(r) e^{i\phi(r)}$, where $f$ is real. The estimate of Proposition \ref{psiconst} yields that $\phi'(r)$ is small for small $d$. Thus, if we change gauge from $(\psi,\hbox{\bf B})$ to $(f,\hbox{\bf Q})$ where $\hbox{\bf Q}=\hbox{\bf B}- \phi'(r)\hbox{\bf e}_r$, then $(f,\hbox{\bf Q})$ is a solution of $(GL_d)$, $\hbox{\bf Q}\cdot \hbox{\bf n}=0$ on $\partial\Omega$  and $|\hbox{\bf Q}|_\infty$ is small. Notice that in this section, we have not explicitely used that $\div\hbox{\bf B} =0$. We have only used that $|\hbox{\bf B}|_\infty$ is small, which was proved in Proposition \ref{aprioriest} under the hypothesis $\div\hbox{\bf B} =0$. Now, since $|\hbox{\bf Q}|_\infty$ is small,  this is enough to do the proof of this section with $(f,\hbox{\bf Q})$ instead of $(\psi,\hbox{\bf B})$. It yields that $\hbox{\bf Q}$ is along $\hbox{\bf e}_\theta$ and that $\phi'=0$. Thus one can choose $\psi$ up to multiplication by a constant of modulus 1. This finishes the proof of Theorem \ref{mainsym}.


\section{Uniqueness for the ball}
In this section, we assume again that $D$ is a ball.
 We have proved in the previous section that we can assume that $\psi=f(r)$ is real and $\hbox{\bf B}=\hbox{\bf Q}=Q(r)\hbox{\bf e}_\theta$. Then the Ginzburg-Landau energy
 in this case is
\beq\label{enradial}
E(f,Q)=\int_D {f'}^2+f^2Q^2+{{\kappa^2d^2}\over 2}(f^2-1)^2+{1\over d^2}({1\over r}(rQ)'-h_0)^2.
\eeq 
We also have $\div \hbox{\bf Q}=0$ then $\|{1\over r}(rQ)'\|_{L^2}$ is an $H^1$ norm. We are going to prove that any radial solution of $(GL_d)$ is a local minimizer of the energy (\ref{enradial}). If $(f,Q)$ is such a solution, then for any $g$ and $\hbox{\bf P}=P(r)\hbox{\bf e}_\theta$ in $H^1$,
\begin{eqnarray}
\nonumber
0&=&
E'(f,Q)\cdot (g,P)
= \int_D f' g'+QPf^2+Q^2fg\\ &+&
{1\over d^2}({1\over r}(rP)')({1\over r}(rQ)'-h_0)+\kappa^2 d^2(f^2-1)fg.\label{en'}
\end{eqnarray}
Moreover
\begin{eqnarray}
\nonumber
E''(f,Q)\cdot (g,P)&=& \int_D {g'}^2+g^2Q^2+4fgQP
+P^2f^2\\ \label{en''}
&+&{1\over d^2}({1\over r}(rP)')^2+\kappa^2 d^2g^2(3f^2-1)fg
\end{eqnarray}

\begin{prop}\label{localmin}
Let $(f,Q)$ be a radial solution of $(GL_d)$ with $d$ and $d\kappa$ small. Then 
 $E''(f,Q)\cdot (g,P) \geq C (\|g/f\|_{H^1}+\| P\|_{H^1})$, where $C$ depends on $d$, $\kappa d$ and $\|f\|_\infty$.
\end{prop}
 The proof follows ideas from \cite{ABG}. First compute $E'(f,Q)\cdot (g^2/f,0)$. This yields
$$
0=\int_D f'\Bigr ( {g^2\over f}\Bigl )'+g^2Q^2+\kappa^2d^2 (f^2-1)g^2.$$
Hence
$$E''(f,Q)\cdot (g,P)= \int_D \Bigr ({\Bigr ({g\over f}\Bigl ) '}\Bigl )^2f^2+4fgQP+
+P^2f^2+{1\over d^2}({1\over r}(rP)')^2+2\kappa^2 d^2g^2f^2.$$
We write
$$4fgQP+\kappa^2 d^2g^2f^2=(\kappa d fg+{2\over {\kappa d}}PQ)^2-{4\over{\kappa^2 d^2}}P^2Q^2.$$
Hence
$$E''(f,Q)\cdot (g,P)\geq \int_D \Bigr ({\Bigr ({g\over f}\Bigl ) '}\Bigl )^2f^2+\kappa^2 d^2g^2f^2+{1\over d^2}({1\over r}(rP)')^2-{4\over{\kappa^2 d^2}}P^2Q^2.$$
Now we use that for $d$ small enough, $\|Q\|_\infty$ is small hence
 $8\int_D P^2 Q^2 \leq \kappa^2 \int_D ({1\over r}(rP)')^2$, thus the estimate  of the Proposition holds.

\hfill

This Proposition yields that any solution of $(GL_d)$ is non degenerate. We are going to prove uniqueness using the bifurcation curve. More precisely, any radial solution of $(GL_d)$ solves
\beq\label{glrad}
\left\{\begin{array}{ll}
f''+{1\over r}f'=f(f^2+Q^2-1)\\
q''+{1\over r}q'-{1\over r^2}q=f^2q,
\end{array}\right.\eeq
with the boundary conditions
\beq\label{bcrad}
f'(0)=f'(1)=0\quad\hbox{and}\quad q(0)=0,\ \Bigr ( q'+{1\over r}q\Bigl ) (1)=h_0.
\eeq
We use shooting techniques, that is solve (\ref{glrad}) with the initial conditions
\beq\label{init}
f(0)=\beta,\ f'(0)=0,\ q(0)=0,\ q'(0)=\alpha.
\eeq
Then the same kind of proof as in Kwong's paper \cite{kwong} which applies the Sturm comparison principle to equation (\ref{glrad}) shows that $f(r;\beta, \alpha)$,  $q(r;\beta, \alpha)$ and $q'(r;\beta, \alpha)$ are increasing functions of $\beta$ and $\alpha$. Moreover, for each $\beta$ in $(0,1)$, there exists a unique $\alpha$ such that $f'(1;\beta, \alpha)=0$, that is the boundary condition is satisfied by $f$. Then
$$h(\beta)=\Bigr (q'+{1\over r}q\Bigl ) (1;\beta, \alpha(\beta))$$
is a continuous function of $\beta$ with limit 0 as $\beta$ tends to 1 and a finite limit as $\beta$ tends to 0. The nondegeneracy result of Proposition \ref{localmin} implies that this curve is decreasing and it provides uniqueness.

Note that uniqueness could also be proved by combining the compactness of solutions and Proposition \ref{localmin} with Crandall Rabinowitz bifurcation Theorem. The Crandall Rabinowitz bifurcation Theorem ensures that only one branch comes off the normal solutions. Then the non degeneracy result of Proposition \ref{localmin} and the implicit function Theorem guarantee that the value $\beta$ continues locally uniquely as $h_0$ is varied. This ensures  uniqueness.


\section{The 1-dimensional case}
\setcounter{equation}{0}
When the superconducting material is an infinite slab of thickness $2d$ between the planes $x=-d$ and $x=d$, it is usual to assume that
 both $\Psi$ and {\bf A} are uniform in the $y$ and $z$ direction, and that
 the exterior magnetic field is
 tangential to the slab, that is $\hbox{\bf H$_0$}=$(0,0,$H_0$).
 A suitable gauge can then be chosen so that $\Psi=f(x)$ is a real function, and 
 $\hbox{\bf A}=q(x)\hbox{\bf e}_y$, where {\bf e}$_y$ is the unit vector
 along the $y$ direction (see \cite{GL} for more details). The model can then be simplified to a system of 2 coupled ODE's $(f,q)$ satisfy
$$\left\{\begin{array}{ll}
{1 \over \kappa^2}f''=f(f^2+q^2-1)
\quad\hbox{in}\quad (-d,d),\\
f'(\pm d)=0,\\
q''=qf^2 \quad\hbox{in}\quad (-d,d),\\
q'(\pm d)=H_0.
\end{array}\right.\eqno{(gl_{d})}$$
In this setting, there are two types of solutions: symmetric solutions where $f$ is even and $q$ is odd, and asymmetric solutions. A complete numerical study of the number and symmetry of these solutions has been done in \cite{AT1}.

Using the same techniques as in the previous sections, we can prove
\begin{theo}\label{theo1d}
There exist constants $d_0$ and $d_1$, such that if $d<\min (d_0,$ $ d_1 / \kappa)$, then any solution of $(gl_{d})$ is symmetric, that is $f$ is even and $q$ is odd.
\end{theo}

This theorem together with the result of uniqueness for symmetric solutions proved in \cite{AT2} give a global uniqueness result for the solutions of $(gl_d)$ with small $d$. 

The proof, as in the previous sections consists first in deriving a priori estimates for the functions $f$ and $q$. We recall from \cite{AT2} that a solution is such that $f$ has a unique maximum, which we call $x_0$, with $\beta= f(x_0)=\|f\|_\infty$, and $q$ is increasing with a unique zero $x_1$. A similar proof to what we did in Section 2 and 3 yields:

\begin{prop}\label{prop1dest}
There exist constants $d_0$ and $d_1$, such that if $d<\min$ $ (d_0, d_1 / \kappa)$, then
\beq\label{est1d}
|{{f(x)}\over \beta} -1|\leq C \kappa d,\ \hbox{and}\
|q(x)-H_0 x|\leq C d^2.
\eeq
\end{prop}

Using similar techniques as in \cite{AT2}, one can get more precise estimates, which in particular give a relation between $\beta$ and $H_0$. 
\begin{prop}\label{prop2dest}
Let $0<d<min(1,1/2\kappa)$, then 
\beq\label{fest}
\beta(1-{{\kappa^2}\over2} (x-x_0)^2)\leq f(x)\leq \beta \quad \forall x\in(-d,d),
\eeq
and there exists $\alpha$ such that
\begin{eqnarray}
\alpha (x-x_1)\leq q(x)\leq \alpha (x-x_1)(1+2\beta d)\quad \forall x\in(x_1,d),\\
\alpha (x-x_1)(1-2\beta d)\leq q(x)\leq \alpha (x-x_1)\quad \forall x\in(-d,x_1),\\
|x_1|\leq 2\kappa^2d^3+2d^2,\\
{\sqrt{3(1-\beta^2)}\over \sqrt{1+3\beta d}}\leq \alpha d \leq {\sqrt{3(1-\beta^2)+2d^2\kappa^2}\over \sqrt{1- d^2\kappa^2/2}}
\end{eqnarray}
\end{prop}
These estimates mean that $f$ is nearly constant and that $q$ is nearly equal to $\alpha x$, because its zero $x_1$ is very small and recall from Proposition \ref{prop1dest} that $\alpha$ and $H_0$ are very close. We will not give the proof of Proposition \ref{prop2dest} since Proposition \ref{prop1dest} is enough for our purposes.

\hfill

The second step of the proof consists in defining the functions $w$ and $z$ as  before (note that the distances have been non dimensionnalized by $d$ as in the previous sections:
\beq\label{wz1d}
w(x)={1\over{\|f\|_\infty}} (f(x)-f(-x))\quad\hbox{and}\quad z(x)={{q(x)+q(-x)}\over d}.
\eeq
We derive the equations satisfied by $w$ and $z$, multiply them respectively by $w$ and $z$ and use the estimates of Proposition \ref{prop1dest} to obtain
$$|w''|\leq \eps|w|+\eps |z|\quad\hbox{and}\quad |z''|\leq \eps|w|+\eps |z|.$$
Then we use the Poincar\'e inequality to get the equivalent of (\ref{wlessz}) and (\ref{zsmall}). 

Finally, we see that since $\int_{-1}^{1} f^2 q=0$, then
\beq\label{final1d}\int_{-1}^{1} z{f^2\over \beta^2}=\int_{-1}^{1} qw{{(\tilde{f}+f)}\over \beta}.\eeq
As previously, we derive that if $z_m$ is different from zero, (\ref{final1d}) provides a contradiction with (\ref{est1d}), which means that $f(x)/\beta$ is nearly constant.

\section{Remarks on the bifurcation curve}
In order to describe the solutions of the Ginzburg-Landau equations, it is common to draw the bifurcation diagram.
 The natural bifurcation diagram, which appears in the physics literature, is to plot the infinity norm of $\Psi$ (that we call $\beta$) against $H_0$. In our setting, recall that $\Psi$ is nearly constant.

For the 1-dimensionnal model, Kwong \cite{kwong} has proved that for each $\beta$, there exists a unique $H_0 (\beta)$ such that $(gl_d)$ has a symmetric solution. Recall that in our setting, we have proved that solutions are indeed symmetric. Aftalion and Troy \cite{AT1} have established a complete picture of the bifurcation curves in the different regimes of $\kappa$ and $d$. In the regime where $d$ is small compared to $1 /\kappa$, then the curve $\beta$ against $H_0$ is decreasing from 1 to 0 as $H_0$ increases from 0 to $C /d$. In \cite{AT2}, it is proved that $\lim_{d\to 0} d H_0 (\beta)=\sqrt{3(1-\beta^2)}$ uniformly with respect to $\beta$ in $(0,1)$. Thus, for $H_0$ larger than $\sqrt 3 /d$, the only solution is the normal solution. For $H_0$ smaller, there exists a unique solution which is not normal; it is symmetric and has $\Psi$ or $f$ almost constant to $\beta$ between 0 and 1. The value of $H_0$ for which the normal solution loses its stability
 has been studied in detail by Bolley-Helffer \cite{BH8}. 

In the 2 dimensionnal case, the same kind of behaviour holds.
 A natural question is to wonder when the superconducting solution exists.
 Using the result of Giorgi-Phillips, we have proved that if there exists a
 non normal solution then $d^2 H_0$ tends to 0 with $d$.
 Hence, if $H_0$ is not small enough, the only solution is the normal 
 solution. On the other hand, it is known that for $H_0$ small, there exists a superconducting solution close to 1.

When there is a superconducting solution, we have also proved that
these solutions are such that $\Psi$ is nearly constant. Let us call
this constant $\beta$. Then there is a relation between the value 
 of $\beta$ and $H_0$. This relation is obtained by integrating
 the equation for $\Psi$ using the boundary condition. One gets
$$\int \Psi (\Psi^2+|A|^2-1)=0.$$
In the 1-d case, since $q(x)$ is equivalent to $H_0 x$, we get
$$\lim_{d\to 0} d^2H_0^2= 3(1-\beta^2),$$
as proved in \cite{AT2}.
In the radial case, one can prove with a similar estimate to Section 2 that $A(r)$ is asymptotically $H_0 r/2$, hence
 $$\lim_{d\to 0} d^2H_0^2= 8(1-\beta^2).$$
This gives an asymptotic limit of the bifurcation curve for small $d$ and finishes the proof of Corollary \ref{coruniq}.

\hfill

\noindent
{\bf Ackowledgements:} The authors would like to thank the referee for helpful suggestions and the second author would like to thank the Ecole Normale Sup\'erieure for their support and hospitality.


\begin{thebibliography}{99}
\bibitem{AT1} {\sc Aftalion, A. \&  Troy, W.C.} 1999 {\em On
 the solutions of the one dimensional Ginzburg-Landau equations for superconductivity.} Physica D {\bf 132} 214-232.
\bibitem{AT2} {\sc Aftalion, A. \&  Troy, W.C.} 1999 {\em Uniqueness
 of solutions of the Ginzburg-Landau equations for thin films.} To appear in EJAM.
\bibitem{ABG}{\sc Alama, S., Bronsard, L., \& Giorgi, T.} 1999 {\em Uniqueness of symmetric vortex solutions in the Ginzburg-Landau
model of superconductivity.} J. Funct. Anal. {\bf 167} 399-424.
\bibitem{BBH} {\sc Bethuel, F., Brezis, H. \& Helein, F.} 1994 {\em Ginzburg-Landau Vortices.}
 Birkh\"auser.
\bibitem{BH8} {\sc Bolley, C. \& Helffer, B.} 1998 {\em Stability of bifurcating solutions for the Ginzburg-Landau equations.} Rev. Math. Phys. {\bf 10} 579-626. 
\bibitem{du}{\sc  Du, Q., Gunzburger, M.D. \& Peterson, J.S.}
 1992 {\em Analysis and
 approximation of the Ginzburg-Landau model of superconductivity.}
 SIAM Review, {\bf 34}, 1, 54-81.
\bibitem{GL} {\sc Ginzburg, V.L. \& Landau, L.D.}
 1950 {\em On the theory of superconductivity.}
 {Soviet Phys. JETP}, {\bf 20}, 1064-1082. English translation: {\em Men of physics: L.D.Landau} ed. D.ter Harr, Pergamon, Oxford (1965) 138-167.
\bibitem{GNN} {\sc Gidas, B., Ni, W.M. \& Nirenberg, L.} 1979 {\em Symmetry and related properties
 via the maximum principle.} Comm. Math. Phys. {\bf 68}, 209-243.
\bibitem{GT} {\sc Gilbarg, D. \& Trudinger, N.S.:} {\em Elliptic partial differential
 equations of second order.} 2nd
edition, Springer Verlag, 1983.
\bibitem{GP} {\sc Giorgi, T. \& Phillips D.} 1999 {\em The breakdown of superconductivity due to strong fields for the Ginzburg-Landau model.} SIAM J. Math. Anal. {\bf 30}, 341-359.
\bibitem{kwong}{\sc Kwong, M.K.}
 1995 {\em On the one-dimensional Ginzburg-Landau BVPs.}
 Diff. Int. Equations {\bf 8}, 1395-1405.
\bibitem{M} {\sc Mironescu, P.} 1996 {\em Les minimiseurs locaux pour l'\'equation de Ginzburg-Landau sont \`a sym\'etrie radiale.} C.R.A.S. Ser.I {\bf 323} 593-598.
\bibitem{SS} {\sc Sandier, E. \& Serfaty, S.} 1999 {\em A rigorous derivation of a free boundary problem arising in superconductivity.} Preprint.
\bibitem{T} {\sc Tinkham, M.} 1975 {\em Introduction to superconductivity.} McGraw-Hill, New-York.
\end{thebibliography}
\end{document}